# A distributed electrical model for superconducting nanowire single photon detectors

Qing-Yuan Zhao[1,2,*], Daniel F. Santavicca[3], Di Zhu[2], Brian Noble[3], and Karl K. Berggren[2,†]

[1]Research Institute of Superconductor Electronics (RISE), School of Electronic Science and Engineering, Nanjing University, Nanjing, Jiangsu 210093, China

[2]Department of Electrical Engineering and Computer Science, Massachusetts Institute of Technology, Cambridge, Massachusetts 02139, USA

[3]Department of Physics, University of North Florida, Jacksonville, Florida 32224, USA

*qyzhao@nju.edu.cn ; [†]berggren@mit.edu

Abstract:

To analyze the switching dynamics and output performance of a superconducting nanowire single photon detector (SNSPD), the nanowire is usually modelled as an inductor in series with a time-varying resistor induced by absorption of a photon. Our recent experimental results show that, due to the effect of kinetic inductance, for a SNSPD made of a nanowire of sufficient length, its geometry length can be comparable to or even longer than the effective wavelength of frequencies contained in the output pulse. In other words, a superconducting nanowire can behave as a distributed transmission line so that the readout pulse depends on the photon detection location and the transmission line properties of the nanowire. Here, we develop a distributed model for a superconducting nanowire and apply it to simulate the output performance of a long nanowire designed into a coplanar waveguide. We compare this coplanar waveguide geometry to a conventional meander nanowire geometry. The simulation results agree well with our experimental observations. With this distributed model, we discussed the importance of microwave design of a nanowire and how impedance matching can affect the output pulse shape. We also discuss how the distributed model affects the growth and decay of the photon-triggered resistive hotspot.

## Article:

Nanostructured superconducting wires are used as a platform for constructing single-photon detectors. A typical superconducting nanowire single-photon detector (SNSPD) is a single long nanowire meandered into a compact two-dimensional shape for efficiently collecting incident photons. Alternative nanowire architectures have been developed to extend the functions of a SNSPD, such as nanowires arranged in parallel for resolving photon numbers[1] or increasing detector output amplitude[2,3] and serial nanowires for photon number resolution[4]. By integrating multiple SNSPDs with on-chip multiplexing circuits, a SNSPD array can also realize spatial sensitivity[5,6].

To understand the detector performance of a particular nanowire geometry, it is important to have an electrical model of the superconducting nanowire. Conventionally, a superconducting nanowire is modelled as an inductor $L_k$ in series with a time-dependent photon-induced resistor $R_n$, as shown in Figure 1a. A numerical electro-thermal model was developed to model the evolution of the resistance growth after a photon detection [7], and a similar phenomenological model was developed based on fitting to experimental observations[8]. These models can simulate the growth of $R_n$, the waveform of the output pulse, and the current recovery in the nanowire. Recently, a model has been developed that integrates the lumped-element model with the SPICE circuit simulation software for convenient implementation in complex circuit systems [9].

However, the lumped element model of a SNSPD is only valid when its geometric length is much shorter than the effective wavelengths in the nanowire. We recently found that effective wavelengths of microwave signals propagating in thin superconducting nanowires are reduced by about two orders of magnitudes from their free space values [10]. This reduction results because a thin superconducting nanowire has significantly higher kinetic inductance than its Faraday inductance. Therefore, a millimeter-long nanowire would introduce ~100 ps propagation delay, which is no longer negligible. For determining the photon detection time, such variations of signal propagation adds additional timing jitter to a SNSPD[11]. These experimental observations suggest that, for nanowire lengths ≈1 mm or longer, it is more accurate to consider a highly inductive superconducting nanowire to be a distributed transmission line and model photon detection as a position-dependent phenomenon.

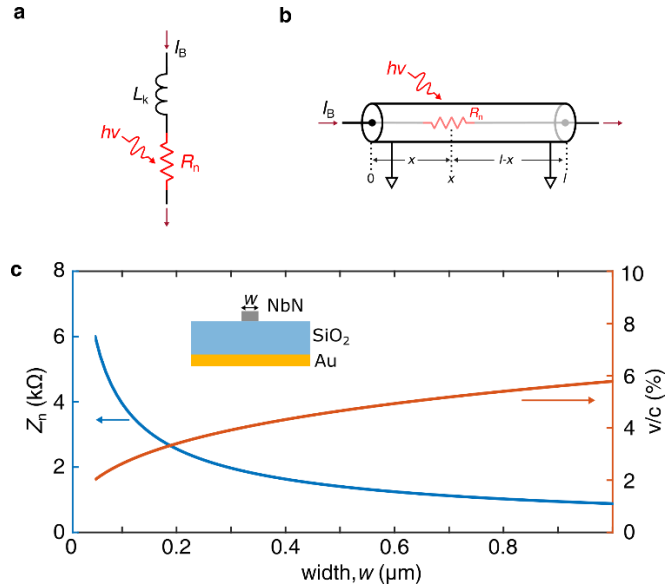

**Figure 1** **(a)** The lumped element model of a highly inductive superconducting nanowire, which is taken as an inductor $L_k$ and a photon-induced resistor $R_n$ in series. The same nanowire can also be modelled as a transmission line, as shown in **(b)**. The total length is $l$. A photon absorption at $x$ destroys the continuity of a superconducting nanowire transmission line and introduces a resistor $R_n$ at $x$. **(c)** Simulated characteristic impedance and velocity for signal propagation along the superconducting transmission line. The simulated transmission line is a 200 μm long straight line. The simulated material stack is air (500 μm)/NbN nanowire (with a sheet inductance of 50 pH per square)/SiO$_2$(300 nm)/Si(500 μm)/perfect conductor.

Therefore, we report a distributed electrical model of a SNSPD and use this model to reproduce output pulses of a long superconducting nanowire patterned into a coplanar waveguide (CPW) structure. As shown in Fig. 1c, the superconducting nanowire is modeled as a transmission line, where a photon detected at location $x$ will trigger the corresponding superconducting nanowire into a resistor $R_n$. As the nanowire is pre-biased with a constant current, such sudden change of resistance will generate a positive pulse propagating from $x$ to 0 and a negative pulse propagating from $x$ to $l$. Compared to the total length of the nanowire, the length of the resistive domain is relatively short. Therefore, the electrothermal model we used for a lumped circuit is still valid for the distributed circuit model to calculate the growth of $R_n$, but the calculation of local current and voltage at $x$ has to consider microwave propagations and reflections based on transmission line theory.

In the distributed model, it is important to extract the characteristic impedance $Z_n$ (which determines the transmission and reflection coefficients) and the signal propagation velocity $v_n$ (which determines the propagation delay) of the equivalent transmission line from the geometry and material properties of a superconducting nanowire and its electromagnetic environment. For a lossless transmission line, the characteristic impedance is $Z_n = \sqrt{L_n/C_n}$ and the phase velocity is $v_n = 1/\sqrt{L_n C_n}$ (where $L_n$ and $C_n$ are the inductance and capacitance per unit length, respectively). The kinetic inductance $L_K$ of the nanowire is much larger than its geometric inductance $L_G$, resulting in $L_n = \frac{(L_K+L_G)}{l} \cong \frac{L_K}{l}$, where $l$ is the total length of the nanowire. $C_n$ is the capacitance per unit length of the nanowire in its surrounding dielectric materials. We estimated $Z_n$ and $v_n$ by simulation using the Sonnet software package, defining a material of sheet inductance of $L_n \times w$. The results are shown in Fig.1c.

For a typical fiber-coupled SNSPD (100 nm wide nanowire, 50% fill factor, 10 μm × 10 μm area), the total length is 0.5 mm, which gives a total maximum delay of 64 ps if the nanowire is considered as a transmission line of $v_n = 0.026 c_0$. This delay is close to the measured jitter of a SNSPD and thus is not easily observed with a single-ended readout [11]. For a longer nanowire, however, transmission line effects become more significant. In figure 2, we show the effect of the meander spacing on the propagation velocity on the nanowire. These simulations were performed using the AXIEM tool in the NI AWR Design Environment software platform. We varied the distance between adjacent lines in the meander (the fill factor) and the total nanowire length. A straight wire with a length of 572 μm was also simulated. The nanowire length was chosen to give approximately the same electrical length for each geometry.

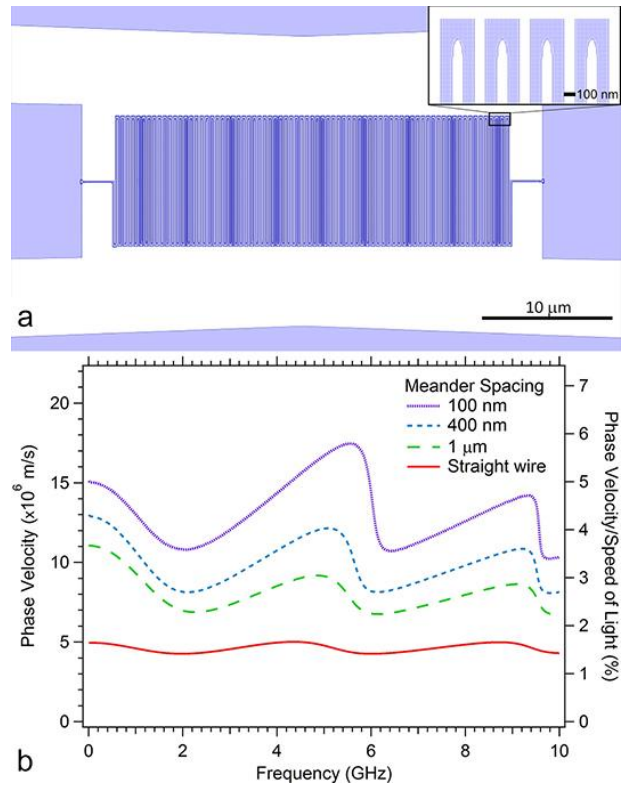

**Figure 2**. (a) Geometry of the nanowire meander (100 nm wide with an inductance of 50 pH/square) with 100 nm meander spacing embedded in the center conductor of a 50 Ω CPW transmission line. Blue is the NbN and white is the sapphire substrate (300 μm thick). (b) Simulated phase velocity as a function of frequency for nanowire meanders with spacings of 100 nm (1.50 mm total length), 400 nm (1.17 mm total length), and 1.0 μm (948 μm total length), along with a straight nanowire (572 μm total length). In all devices, the nanowire width is 100 nm. The nanowire length for each geometry was chosen to have a first-order self-resonance at approximately 5 GHz.

We see that the phase velocity, and hence also the characteristic impedance, vary significantly as a function of frequency for the meander geometry. The more tightly packed the meander, the faster the phase velocity. For a given nanowire length, this effect results in a decreased ability to determine the photon absorption location based on a comparison of the timing of the output pulses from each end of the nanowire. The periodic structure seen in the phase velocity as a function of frequency results from standing wave resonances along the length of the nanowire[10]; each peak in the propagation velocity corresponds to a half-wave resonance, with the first order (λ/2) resonance at approximately 5 GHz.

The frequency-dependent phase velocity exhibited by the meander geometry results in pulse dispersion. To illustrate this, we show in figure 3a a comparison of the rising edge of output pulses due to photon-induced hotspots in three different locations along the nanowire, one at each end and a third in the middle. A clear time separation of the pulses is only observed at the very beginning of the rise, and this time separation is relatively small. In a real device, noise would likely prevent triggering at a sufficiently low voltage to capture this time separation. We conclude that accurately resolving the photon detection location based on pulse timing would not be practical for this nanowire geometry. This pulse dispersion will also result in the geometrical contribution to the measured device jitter varying based on the trigger threshold voltage.

Figure 3b shows the initial pulse rise for a straight nanowire with a length of 0.5 mm, which corresponds to an equivalent electrical length to the longer meander. The structure in the pulse rise shown in figure 3b is due to the large impedance mismatch between the nanowire and the 50 Ω readout. The pulse timing, and hence the location of photon absorption, can still be accurately determined from the initial sharp, well-defined portion of the pulse rising edge, which is about 50% of the total pulse height. An impedance taper could be used to impedance-match the higher-frequency pulse components, eliminating most of this structure from the rising edge [12].

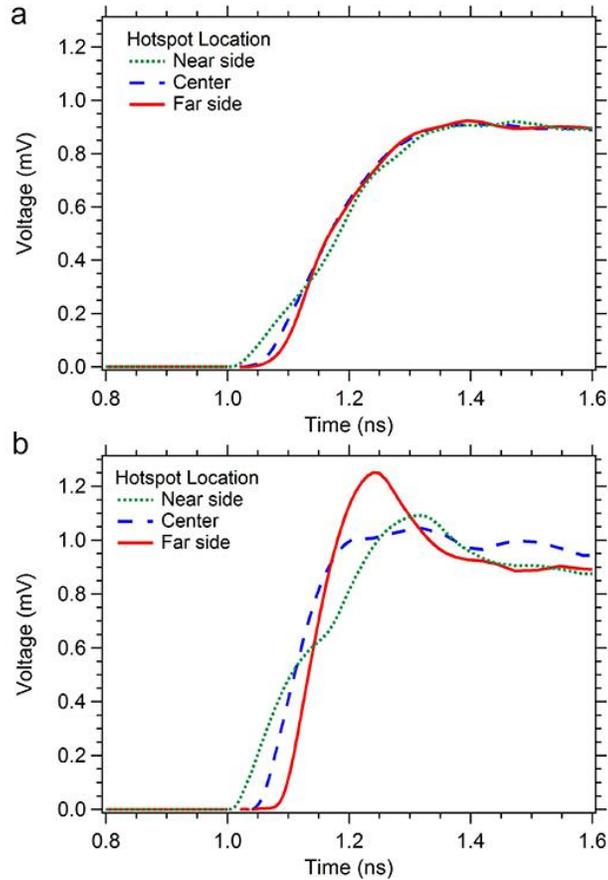

**Figure 3** Simulation of rising edge of pulses measured at one end of the nanowire that are produced by a 6 kΩ resistive hotspot of 300 ps duration. The sheet inductance is set to 50 pH/square and the bias current is 20 µA. Pulses are simulated for hotspots at each end of the nanowire, as well as in the center. Pulses are passed through a 5 GHz low pass filter to match the bandwidth of a typical SNSPD readout. (a) Results for a 100 nm wide nanowire patterned in a meander with 100 nm meander spacing (50% fill factor) and total length of 1.5 mm. (b) Results for a straight 100 nm wide nanowire with a total length of 0.5 mm, which is approximately the same electrical length as the 1.5 mm long meander.

Previous simulation results of a SNSPD suggest that if a clear dependence of detection pulses on the photon absorption locations is wanted, we should design the device with minimal dispersion and read it with an impedance matched readout. Such a device was recently demonstrated as a single-photon imager [12]. In this device, the nanowire was patterned into a CPW structure with a characteristic impedance of 1.0 kΩ and a signal velocity of 5.56 μm/ps. The total length of the nanowire was 19.7 mm. This resulted in a maximum propagation delay of ≈ 3.2 ns, a significant delay that we can measure precisely with a 6 GHz bandwidth oscilloscope. To minimize impedance mismatch between the nanowire and the readout circuit, we designed the ends of the nanowire into Klopfenstein tapers to transform the impedance from 1.0 kΩ to 50 Ω [13]. The taper had a passband starting from 700 MHz, ensuring the rising edge of the photon detection pulse would be efficiently transmitted. As a photon detection triggered two electrical pulses of opposite propagation directions toward to each end of the nanowire, we used a timing-differential readout to collect both of them.

As the nanowire was designed into a CPW, dispersion was assumed to be negligible. Thus, a lossless transmission line model in SPICE can be used for representing the pulse propagation along the nanowire. As shown in Fig. 4a, we inserted a hotspot at position $x$ in a transmission line that has a length of $l$. To simulate the dynamics of the hotspot resistor $R_n$, we used a SPICE model of a superconducting nanowire [9]. This model simplified the calculation of $R_n$ with adequate accuracy for estimating the switching effects in an SNSPD. We simulated a distributed circuit of a conventional meandered SNSPD in SPICE, which gave similar shapes of the rising edges shown in Figure 3 but reduced the simulation time to a few seconds. To model the tapered nanowire for impedance matching, we divided the taper into 100 transmission lines of incrementally varying impedance and connected them in series.

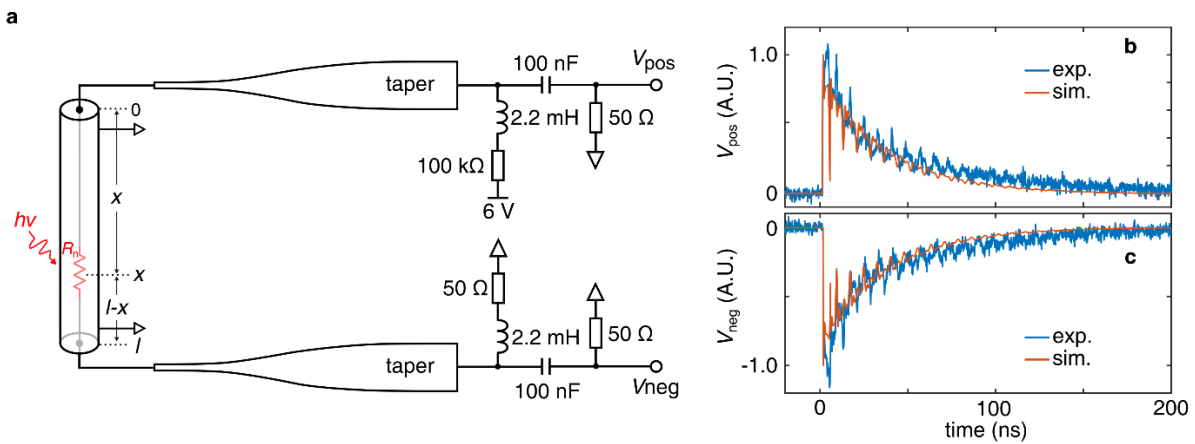

**Figure 4**. Simulated output pulses from a nanowire designed into a CPW with an impedance matched readout. **(a)** Schematic diagram of the simulated circuit. Each end of the nanowire connects to a tapered nanowire for transforming the impedance. The readout circuit reads out the output pulses from the two ends of the nanowire, which are shown in **(b)** and **(c)**. In (b) and (c), the blue traces are from experimental acquisition while the red traces are from the simulation. The firing location is at $x = 8278$ μm. The amplitudes of these pulses are normalized to the first maximum.

As shown in Fig. 3b and c, the distributed model can reproduce similar pulse shapes to those seen experimentally. In particular, the pulses have similar reflection ripples on the falling edge. Because the impedance transform taper has a low-cutoff frequency of 700 MHz, a lumped description is still valid for low frequency signals. Thus, the pulse envelope follows an exponential decay. In the simulation, the total inductance of the nanowire was set to $L_t = 3.2$ μH, giving a decay constant of $L_t/(2 \times 50\ \Omega) = 32$ ns.

The more prominent features of the distributed nanowire model were exhibited at the first transmitted pulses, which are shown in Fig. 4. We simulated two different firing locations ($x = 1668$ μm and $x = 8278$ μm) and compared the output traces to our experimental observations. Both the simulation pulses and experimental pulses showed that the rising edges of the output pulses were well preserved so that the pulse arrival times were linearly proportional to $x/v$. Although the nanowire had a total inductance as large as 3.2 μH, the rising time of the edge was about 100 ps for the simulation pulses and ~250 ps for the experimental pulses. It indicates that the rising edge of the pulses from the distributed detector reflected the transition time from a superconducting state to a normal resistor, but not the overall lifetime of the normal resistor. The simulated pulses succeeded in producing ripples of large amplitude. However, some small ripples did not match to our experimental results, which was probably because the distributed model did not include impedance mismatch from bonding wires and connections and other parasitic parameters.

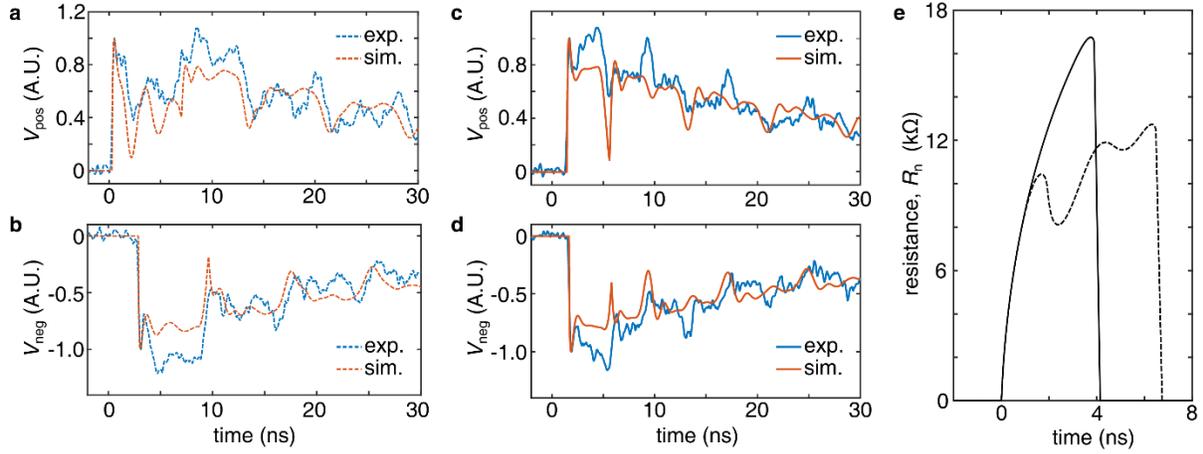

**Figure 5.** Details of the simulated (red) and experimental (blue) pulses at two firing locations. (**a**) and (**b**) are the two output pulses for $x = 1668$ μm, while (**c**) and (**d**) are the two output pulses for $x = 8278$ μm. (**e**) Simulated growth of the normal resistance $R_n$ for the two firing cases (dashed line for $x = 1668$ μm and solid line for $x = 8278$ μm).

The distributed model indicates that the growth of the normal region $R_n$ depends on the photon detection location. As shown in Fig. 5e, when a photon was detected in the middle ($x = 8278$ μm), $R_n$ reached a maximum value of 17 kΩ within a lifetime of 4.1 ns. For the detection event at the end close to the readout amplifier ($x = 1668$ μm), $R_n$ increased to a maximum value of 13 kΩ after oscillating a few times. The lifetime of the normal domain increased to 6.7 ns. The difference of growth of $R_n$ can be explained by analyzing the pulse reflections in the distributed model. If the photon detection happened at the middle of the nanowire, the hotspot kept growing until the two reflected pulses returned to the detection region. Because the nanowire impedance was larger than the load impedance, the pulses flipped into reversed pulses (in direction and amplitude) and reflected back to $R_n$. Since $x$ was at the middle of the nanowire, these reflected pulses arrived at $R_n$ at almost the same time. This event resulted in a sudden reduction of the local current through $R_n$ to below the level for sustaining a normal domain. Consequently, the normal domain was reset to the superconducting state. For the case where $x = 1668$ μm, the pulse propagating to the near end reflected back to the normal domain earlier than the other pulse propagating to the far end. The first reflected pulse decreased the local current to reduce $R_n$, but this reduction was not enough to bring $R_n$ to zero. A portion of the pulse then returned to the near-end, repeating the process. The red curve in Fig. 5c shows two dips, corresponding to the reduction of current causing by two reflected pulses. Finally, the original detection pulse propagating to the far-end of nanowire reflected to $R_n$, adding another reduction of the local current, the normal domain can reset to the superconducting state.

In conclusion, we created a distributed model for a superconducting nanowire and applied this model to study the photon response pulses from a superconducting nanowire with different geometries and readout schemes. Our simulation results suggest that, to observe a clear position dependence of the output pulses on photon detection locations, the microwave design of the nanowire should be considered. The distributed model also showed that the growth of the hotspot resistance depended on its locations, which could be explained by considering pulse propagations and reflections.

With this model, we can design a distributed nanowire detector and use the microwave readout to extract additional detection information, *e.g.* to map the photon detection locations from the pulse arrival times[12] or to resolve multi-photon detection events in a serial detector array[14]. The distributed model can also be used in frequency multiplexed detector arrays to simulate the transient pulses[15]. We envision this distributed model will be a useful electrical tool for developing new nanowire-based devices and microwave circuits.

## Acknowledgments

This research was supported by the National Science Foundation (NSF) grants under contact No. ECCS-1509486 (MIT) and No. ECCS-1509253 (UNF), and the Air Force Office of Scientific Research (AFOSR) grant under contract No. FA9550-14-1-0052. Di Zhu is supported by National Science Scholarship from A*STAR, Singapore. Qing-Yuan Zhao is partially supported by the Fundamental Research Funds for the Central Universities No.021014380100.